\documentclass{aastex63}

\usepackage{amsmath}
\usepackage{amssymb}
\usepackage{graphicx}
\usepackage{CJKutf8}
\begin{document}
\begin{CJK}{UTF8}{gbsn}

\title{Three-dimensional Turbulent Reconnection within Solar Flare Current Sheet}

\correspondingauthor{Xin Cheng}
\email{xincheng@nju.edu.cn}

\author[0000-0001-9863-5917]{Yulei Wang}
\affiliation{School of Astronomy and Space Science, Nanjing University, Nanjing 210023, People's Republic of China}
\affiliation{Key Laboratory for Modern Astronomy and Astrophysics (Nanjing University), Ministry of Education, Nanjing 210023, People's Republic of China}

\author[0000-0003-2837-7136]{Xin Cheng}
\affiliation{School of Astronomy and Space Science, Nanjing University, Nanjing 210023, People's Republic of China}
\affiliation{Key Laboratory for Modern Astronomy and Astrophysics (Nanjing University), Ministry of Education, Nanjing 210023, People's Republic of China}

\author[0000-0002-4978-4972]{Mingde Ding}
\affiliation{School of Astronomy and Space Science, Nanjing University, Nanjing 210023, People's Republic of China}
\affiliation{Key Laboratory for Modern Astronomy and Astrophysics (Nanjing University), Ministry of Education, Nanjing 210023, People's Republic of China}

\author{Zhaoyuan Liu}
\affiliation{Shandong Computer Science Center (National Supercomputing Center in Jinan), Qilu University of Technology (Shandong Academy of Sciences), Jinan 250014, People's Republic of China}
\affiliation{Advanced Algorithm Joint Lab, Shandong Computer Science Center, Qilu University of Technology, Jinan 250014, People's Republic of China}

\author[0000-0001-7484-401X]{Jian Liu}
\affiliation{Advanced Algorithm Joint Lab, Shandong Computer Science Center, Qilu University of Technology, Jinan 250014, People's Republic of China}
\affiliation{School of Nuclear Science and Technology, University of Science and Technology of China, Hefei 230026, People's Republic of China}

\author[0000-0002-7878-0655]{Xiaojue Zhu}
\affiliation{Max Planck Institute for Solar System Research, Gottingen, D-37077, Germany}

\begin{abstract}
Solar flares can release coronal magnetic energy explosively and may impact the safety of near-earth space environments.
Their structures and properties on macroscale have been interpreted successfully by the generally-accepted two-dimension standard model invoking magnetic reconnection theory as the key energy conversion mechanism.
Nevertheless, some momentous dynamical features as discovered by recent high-resolution observations remain elusive.
Here, we report a self-consistent high-resolution three-dimension magnetohydrodynamical simulation of turbulent magnetic reconnection within a flare current sheet.
It is found that fragmented current patches of different scales are spontaneously generated with a well-developed turbulence spectrum at the current sheet, as well as at the flare loop-top region.
The close coupling of tearing-mode and Kelvin-Helmholtz instabilities plays a critical role in developing turbulent reconnection and in forming dynamical structures with synthetic observables in good agreement with realistic observations.
The sophisticated modeling makes a paradigm shift from the traditional to three-dimension turbulent reconnection model unifying flare dynamical structures of different scales.
\end{abstract}

\section{Introduction}

Various explosive celestial phenomena including solar flares \citep{Masuda1994}, supernova remnant shocks \citep{Matsumoto2015}, and quasar jets \citep{Shukla2020} often invoke magnetic reconnection as a vital energy release mechanism.
As the first reconnection theory, the Sweet-Parker (SP) model predicts a stationary reconnection rate within the framework of two-dimensional (2D) current sheet (CS) of a large aspect ratio \citep{Sweet1958}.
Nevertheless, the corresponding reconnection rate is too low to interpret explosive energy release, especially for those in low-resistivity plasma environment \citep{Priest2000}.
As the inheritor of the SP model, the Petscheck model significantly shortens the CS and introduces two pairs of slow-mode shocks at its two ends \citep{Petschek1964}.
The reconnection rate is largely boosted and well comparable with observations.
Over the last decades, with a number of observations and simulations, a Petscheck-type 2D standard flare model (CSHKP model) that is capable of explaining various macroscale flare features has been established \citep{Carmichael1964,Sturrock1966,Hirayama1974,Kopp1976,Shibata1995,Lin2000}.

Theoretically, the magnetic reconnection process is believed to be highly dynamic in essence.
The elongated CS could be fragmented into many pieces of different scales by the tearing mode instability (TMI) \citep{Loureiro2007,Bhattacharjee2009}, which can modulate the reconnection rate effectively and even make the dissipation region highly turbulent \citep{Dong2022}.
If including the third dimension, the TMI can produce oblique turbulent eddies distributed along deformed magnetic field lines \citep{Huang2016,Beresnyak2017,Leake2020}.
The Kelvin-Helmholtz instability (KHI) may appear and grow in the later stage of magnetic reconnection  \citep{Kowal2017,Kowal2020}, giving rise to the stochastic magnetic field, which can further facilitate the fast reconnection \citep{Lazarian1999,Kowal2009}.
However, if the guide field along the third dimension is too strong, the 3D reconnection may be similar to the 2D case, showing weakly deformed flux ropes \citep{Daldorff2022}.

The fine structures/processes in the reconnection CS could be closely related to various small-scale flare structures.
Bright plasma blobs are often observed to appear in the long-stretched flare CS \citep{Asai2004,Nishizuka2009,Takasao2016a}, indicating the appearance of TMI-induced plasmoids as confirmed by 2D numerical models \citep{Shen2011,Barta2011} and laser experiments \citep{Ping2023}.
Different from 2D results, the CS in three-dimension (3D) simulations is suggested to be highly fragmented \citep{Nishida2013,Edmondson2017}, which could be promising for explaining super-hot turbulent plasma within the CS \citep{warren2018}, supra-arcade downflows (SADs) and finger-like supra-arcade fans (SAFs) above the flare loop top \citep{McKenzie1999a,McKenzie2013,Savage2012}.
On the other hand, the reconnection-generated downflows will collide with the dense plasma at the flare loop top, even forming a termination shock \citep{Chen2015,Shen2018a}.
Due to the non-uniformity of magnetic reconnection caused by the CS fragmentation, the downflows are intermittent with varying velocity \citep{Cheng2018}, which tend to cause a highly turbulent region between the termination shock and the flare loop top \citep{Ye2020,Shibata2023}.

Although the 3D patchy CS structures have been numerically documented \citep{Linton2009,Nishida2013,Edmondson2017}, they were implemented in ideal configurations, thus preventing from connecting small-scale reconnection processes in the CS to observed flare fine structures, in particular above the flare loop top.
\cite{Shen2022} and \cite{Ruan2023} analyzed the instabilities and turbulence at the flare loop top but did not explore their CS origin that could be of the first principle.
Up to now, the whole causal chain of 3D turbulent reconnection in a realistic flare configuration, including the growth of instabilities, the spontaneous formation of flux ropes (plasmoids in 2D), the development of turbulence in the CS, the outflow-induced turbulence at the flare loop top, and the sub-structures at the flare ribbon fronts, has not been realized in a self-consistent magnetohydrodynamical (MHD) simulation.
Higher resolution and necessary flare physics are indispensable for such a simulation.
On the one hand, the numerical resistivity can be substantially reduced to realize the physics during small-scale reconnection processes; on the other hand, a direct comparison with flare observations is possible.

\section{Numerical Model}

In this paper, we perform an MHD simulation of 3D magnetic reconnection in a flare CS and implement an unprecedented uniform spatial resolution of $26\,\mathrm{km}$ in the key regions enclosing the CS and loop top (20\% of total domain with a length-scale up to $100\,\mathrm{Mm}$).
Key coronal effects, i.e., gravity, anisotropic thermal condition, radiative cooling, and background heating, are included in the resistive MHD equations:
\begin{eqnarray}
\frac{\partial\rho}{\partial t}+\nabla\cdot\left(\rho{\bf u}\right) & = & 0\,,\nonumber \\
\frac{\partial\left(\rho{\bf u}\right)}{\partial t}+\nabla\cdot\left(\rho{\bf u}{\bf u}-{\bf BB}+P^{*}{\bf I}\right) & = & \rho\bf{g}\,,\nonumber \\
\frac{\partial e}{\partial t}+\nabla\cdot\left[\left(e+P^{*}\right){\bf u}-{\bf B}\left(\bf{B}\cdot{\bf u}\right)\right] & = & \rho{\bf g}\cdot {\bf u}+\nabla\cdot\left(\kappa_{\parallel}\hat{\bf{b}}\hat{\bf{b}}\cdot\nabla T\right)+C_{rc}\left(H-R\right)\,,\label{eq:MHD}\\
\frac{\partial{\bf B}}{\partial t}-\nabla\times\left(\bf{u}\times\bf{B}\right) & = & -\nabla\times\left(\eta{\bf J}\right)\,,\nonumber \\
{\bf J} & = & \nabla\times{\bf B}\,,\nonumber
\end{eqnarray}
where the adiabatic index is $\gamma=5/3$ and standard notations of variables are used.
The gravity acceleration is evaluated by ${\bf g}=-g\hat{\bf{e}}_y$, where $g=g_0/\left(1+y/R_{\odot}\right)^2$, $R_{\odot}$ denotes the solar radii, and $g_0=2.7390\times 10^4\,\mathrm{cm\,s^{-2}}$ is the gravity acceleration at the solar surface.
The conductivity parallel with magnetic field is determined by $\kappa_\parallel=\kappa_0T^{2.5}$, where $\kappa_0=5\times10^{-7}\,\mathrm{erg\,s^{-1}\,cm^{-1}\,K^{-3.5}}$.
The optically thin radiative cooling term $R=n_en_i\Lambda\left(T\right)$ is based on a widely used model in which $n_e$ and $n_i$ are respectively the number density of electrons and ions and $\Lambda\left(T\right)$ is a piece-wise cooling function \citep{Klimchuk2008}.
A temporally static background heating is set as $H=n_in_e\Lambda\left(T_\mathrm{init}\right)$ to balance the cooling at the initial state, where $T_\mathrm{init}$ is the initial temperature distribution.
To avoid the numerical instabilities at the bottom region, we have multiplied $H$ and $R$ by a dimensionless index $C_{rc}=0.5\mathrm{tanh}\left[\left(y-0.3\right)/0.03\right]+0.5$ to eliminate the effects of cooling and heating in the region below $y=0.2$.
In our simulation, all physical quantities are normalized based on the same dimensionless units as \cite{Wang2022}.

The initial conditions of temperature, pressure, and magnetic field are the same as \cite{Wang2022}.
To trigger the fast reconnection, we set a temporally damping localized anomalous resistivity on the background low-resistivity plasmas, namely,
\begin{equation}
\eta=\eta_b+\eta_{a}\mathrm{exp}\left[-\frac{x^{2}+\left(y-h_{a}\right)^{2}}{l_{a}^{2}}\right]\mathrm{exp}\left(-\frac{t^2}{t_a^2}\right)\,,\label{eq:eta}
\end{equation}
where, $\eta_a=1\times10^{-3}$, $h_a=0.5$, $l_a=0.03$, $t_a=2$, and the background resistivity is uniformly set as $\eta_b=5\times10^{-6}$, which corresponds to a Lundquist number $S=L_0u_0/\eta_b=2\times 10^5$.
The anomalous resistivity only works at the initial stage.
After $t=5$, the effect of $\eta_a$ is ignorable and the evolution is dominated by the background resistivity $\eta_b$.
The initial background velocity is set as zero everywhere.
To trigger the self-sustained 3D turbulence \citep{Huang2016}, in the CS region, we add a small-amplitude Gaussian thermal noise of velocity with a zero mean value and a standard deviation $\sigma_u=10^{-3}$.

The simulation domain is set as $x\in[-0.5,0.5]$, $y\in[0,2]$, and $z\in[-0.15,0.15]$.
The bottom ($y=0$) is a symmetric boundary, the top ($y=2$) is a no-inflow boundary, and the rest are open boundaries \citep{Yokoyama2001,Wang2021b,Kou2022}.
We only use the data in $\lvert x\rvert<0.3$ and $y<1$ for analysis in order to reduce the influences of numerical boundaries.
Compared with adaptive mesh refinement (AMR), the uniform spatial mesh is a better choice for the direct numerical simulations (DNSs) of turbulence, which can avoid numerical perturbations brought by mesh refining and provide uniform resolution in turbulent regions \citep{Shibata2023}.
We use the static mesh refinement (SMR) technique to implement a uniform mesh in the core reconnection region and also save computational costs.
We set three refinement levels.
The root level-0 grid numbers are $480$, $960$, and $144$ on $x$, $y$, and $z$ directions, respectively.
In the domain $0.1<\lvert x\rvert<0.3$, the level-1 grid number doubles in three directions.
In the core turbulent region, $\lvert x\rvert\leq 0.1$, the level-2 grid refines again, which achieves the highest resolution equivalent to an effective mesh of $1920\times 3840\times 576$.
The pixel scale there is $\Delta L=\Delta x=\Delta y=\Delta z=26\,\mathrm{km}$.

We numerically solve the above system using the \textsf{Athena++} code \citep{Stone2020}.
The conservation part of the MHD equation is solved by the HLLD Riemann solver \citep{Miyoshi2005}, the 2-order piecewise linear method (PLM), and the 2-order van Leer predictor-corrector scheme.
The resistivity, thermal conduction, gravity, radiative cooling, and heating terms are calculated by the explicit operator splitting method.
We use the 2-order RKL2 super-time-stepping algorithm to reduce computational costs \citep{Meyer2014}.
The simulation stops at $t=8.2$, which corresponds to $15.66$ minutes in physical time ($t_0=114.61\,\mathrm{s}$).

\begin{figure*}[h]
\gridline{\fig{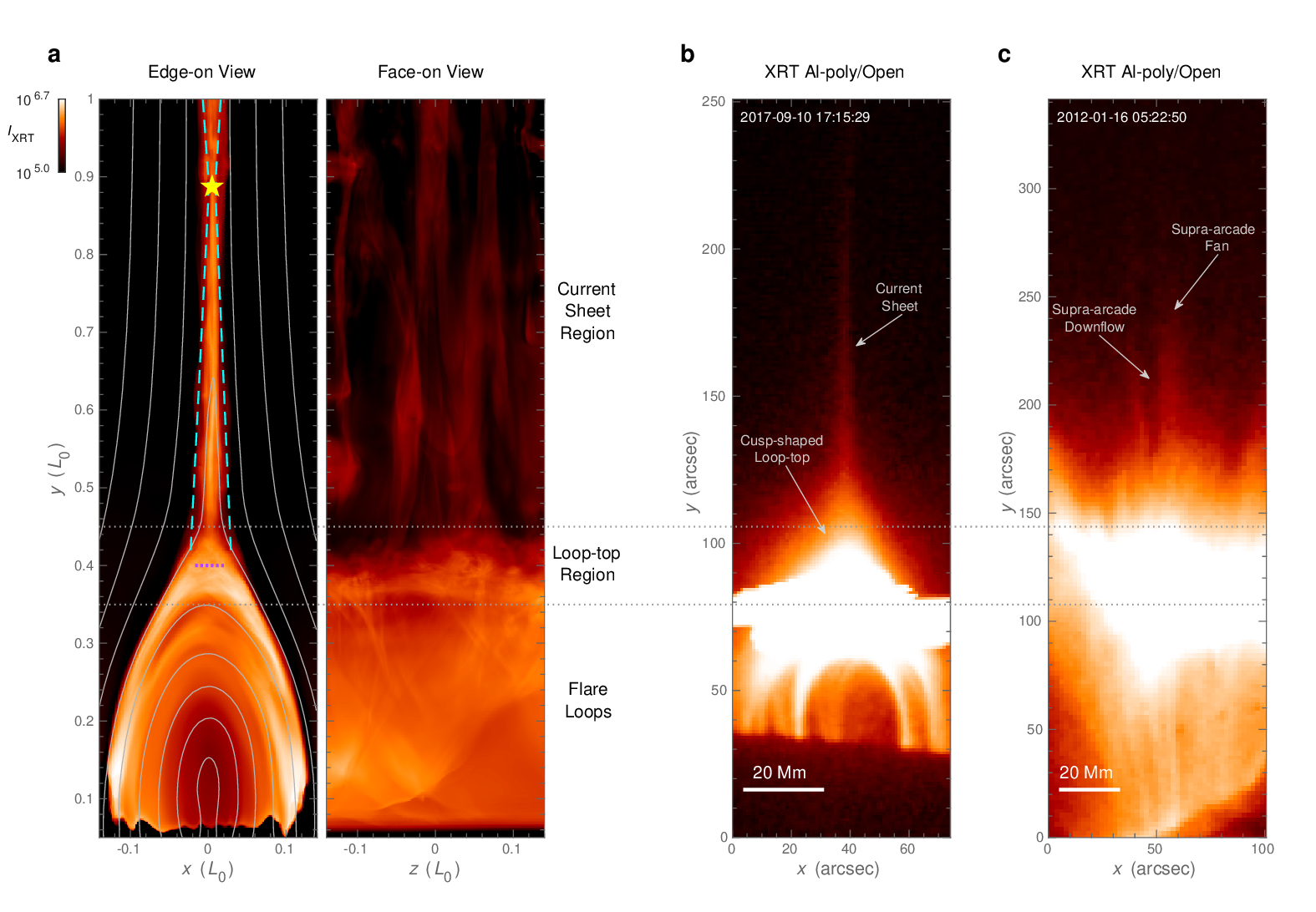}{1\textwidth}{}} 
\caption{Observational characteristics of flare turbulent reconnection.
(a) Synthesized XRT Al-poly/Open images of the edge-on (left panel) and face-on (right panel) CS  and flare loops at $t=8.2$.
In the left panel, the grey curves depict the $z$-averaged in-plane magnetic field, the yellow star represents the principal reconnection point of the mean magnetic field, and the cyan and purple dashed lines schematically plot the positions of slow-shocks and the loop-top termination shock predicted by the standard flare model, respectively.
The principal reconnection point is identified as the position with the maximum $z$-averaged out-of-plane current density $\bar{J}_z$ \citep{Wang2021b}.
The magenta and green dashed lines in the face-on image mark the positions of BS and DS in Fig.\,\ref{fig4}, respectively.
The units of length is $L_0=5\times 10^9\,\mathrm{cm}$.
(b) The edge-on view of the reconnecting CS and flare loops at 17:15:29 UT on 2017 September 10 observed by the XRT.
(c) The face-on view of the flare at 05:22:50 UT on 2012 January 01 observed by the XRT.
\label{fig1}}
\end{figure*}

\section{Results}
\subsection{Flare Turbulent Reconnection}

To compare with observations, we use the simulated density and temperature to synthesize observational images.
As we suppose an optically thin atmosphere, the intensity is evaluated by $I=\int n_e^2f\left(T\right)\mathrm{d}l$, where $f\left(T\right)$ is the temperature response function of instruments and $\mathrm{d}l$ is the line element along the line of sight (LOS) \citep[also see][]{Zhao2019,Ye2020,Xie2022,Shen2022,Ruan2023}.
When synthesizing the edge-on images, we set the domain of LOS as $z\in\left[-0.14,0.14\right]$.
For the face-on images, the LOS domain is $x\in\left[-0.14,0.14\right]$, which encloses the flare loop region.
The depths of LOS for both views are the same for the sake of a better comparison.

At the end of our simulation, in both the CS and loop top, fully-developed turbulence is finally achieved and key observational features are self-consistently generated (Fig.\,\ref{fig1}).
The edge-on view of the synthesized XRT image reveals a well-known configuration complying with the standard flare model, including flare loops, cusp-shaped loop top, and stretched CS (Fig.\,\ref{fig1}a, b).
The flare loops are relatively stable as reinforced by the downward reconnected flux, while the loop-top and the CS regions are highly turbulent.
As observed from the face-on view, the CS region presents alternately dark and bright structures, highly resembling bright SAFs and dark SADs in observations (Fig.\,\ref{fig1}a, c).

\begin{figure*}[h]
\gridline{\fig{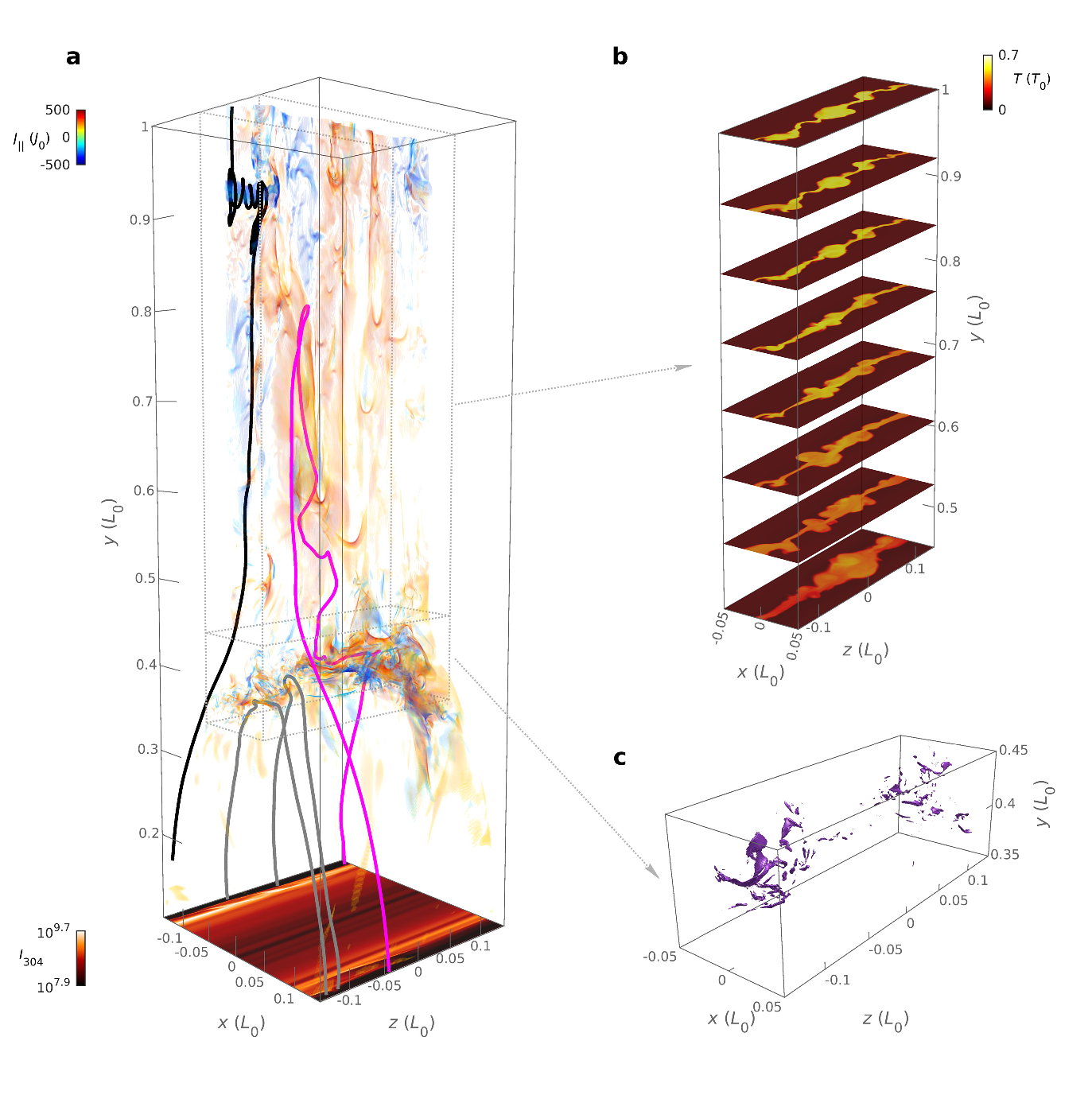}{0.9\textwidth}{}} 
\caption{3D structures of flare turbulent reconnection.
(a) The profile of strong parallel current density ($\left|J_{\parallel}\right|>150$) at $t=8.2$.
The black and magenta curves depict two typical magnetic field lines, whose upper part wraps strong current patches (flux ropes) in the turbulent CS but footpoints away from the flare ribbon.
The grey curves show the magnetic field lines from the front of the flare ribbons crossing the turbulent structure in the loop-top region.
The frame bottom ($y=0.1$) overlapped with an AIA 304 \AA\ image synthesized from the top view with a LOS ranging $y\in\left[0,1\right]$, which shows two parallel flare ribbons at the lower atmosphere.
The two dotted boxes highlight the upper CS and the lower loop-top regions, respectively.
(b) The distributions of the CS temperature in the $x$-$z$ planes at different heights.
(c) The distribution of the termination shocks at the loop-top region.
The units of current density and temperature are $J_0=9.54\,\mathrm{statC\,s^{-1}\,cm^{-2}}$ and $T_0=1.15\times 10^{7}\,\mathrm{K}$, respectively.
An animation of this figure showing the evolution from $t=5$ to $8.2$ is available.
\label{fig2}}
\end{figure*}

Our results are consistent with the Petscheck-type standard model on the macro level that contains a principal X-point, two slow-mode shocks enclosing reconnection outflows, and a termination shock at the loop top (Fig.\,\ref{fig1}a).
However, a number of new fine structures are revealed within reconnecting CS and flare loop top (Fig.\,\ref{fig2}).
Because of the background resistivity $\eta_b$ being uniform, the magnetic dissipation regions can be represented by the profile of strong parallel current density, $J_\parallel=E_\parallel/\eta$, where $E_\parallel$ is the electric field parallel with local magnetic field \citep{Reid2020}.
The CS is fragmented into current patches of various shapes and scales, including several relatively large-scale curled current patches that grow vertically (see the region passed by the magenta field line in Fig.\,\ref{fig2}a), some horizontal flux ropes (see the horizontal negative current tube wrapped by the black curve in Fig.\,\ref{fig2}a), and many small-scale current patches that represent distorted flux ropes (Fig.\,\ref{fig2}a).
The morphology of the reconnection structures varying in spatial scales is largely different from the results in 2D simulations \citep{Shen2011,Barta2011}, as well as the prediction of 2D fractal reconnection model \citep{Shibata2001}.

\subsection{Turbulence Development}

\begin{figure*}[h]
\gridline{\fig{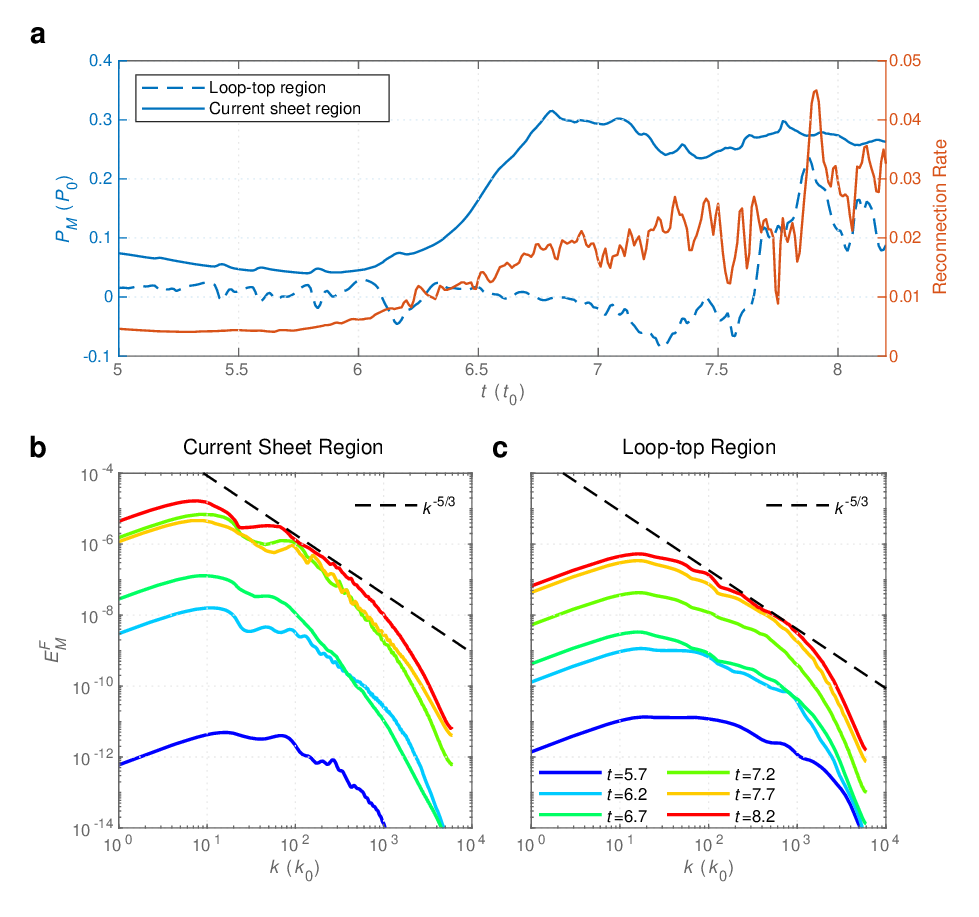}{0.7\textwidth}{}} 
\caption{Temporal evolution of turbulent reconnection at the CS and loop top.
(a) The temporal evolution of volume-averaged effective magnetic energy release rates at the CS (blue solid curve) and loop top (blue dashed curve).
The orange curve displays the evolution of the reconnection rate.
(b and c) The power spectra of turbulent magnetic energy at the CS and loop top at selected moments as denoted by different colors. The oblique dashed line denotes the classic spectral index of $-5/3$ as predicted in theory.
\label{fig3}}
\end{figure*}

In the CS region, the development of turbulent reconnection is divided into three stages.
During the precursor stage, $0\leq t<5.7$, the reconnection presents a moderate rate (Fig.\,\ref{fig3}a) and basically follows the behaviors as found in 2D simulations, i.e., the CS gets thinner, the flare loops gradually form, and horizontal flux ropes (plasmoids in 2D) are generated by the TMI (see the Movie of Fig.\,\ref{fig2}).
Because the 3D reconnection rate is difficult to calculate precisely, here we estimate it by the 2D reconnection rate of the mean magnetic field $\bar{\bf B}=\left<\bf B\right>_z$, where $\left<\cdot\right>_z$ means taking the average value on the entire $z$ direction \citep{Huang2016}.
We first calculate the flux function $\psi\left(y,t\right)\equiv\int_0^y\bar{B}_x\left(y',t\right)\mathrm{d}y'$ along the mid-plane ($x=0$).
Then, the reconnected flux at a moment $t$ is determined by $\Psi\left(t\right)=\mathrm{max}\left(\psi\right)-\mathrm{min}\left(\psi\right)$.
Finally, the reconnection rate is calculated by $\partial_t{\Psi}/B_0C_{A0}$, where $C_{A0}=B_0/\sqrt{\rho_0}$ is the background Alfv\'{e}n speed \citep[also see][]{Zenitani2020}.

\begin{figure*}[h]
\gridline{\fig{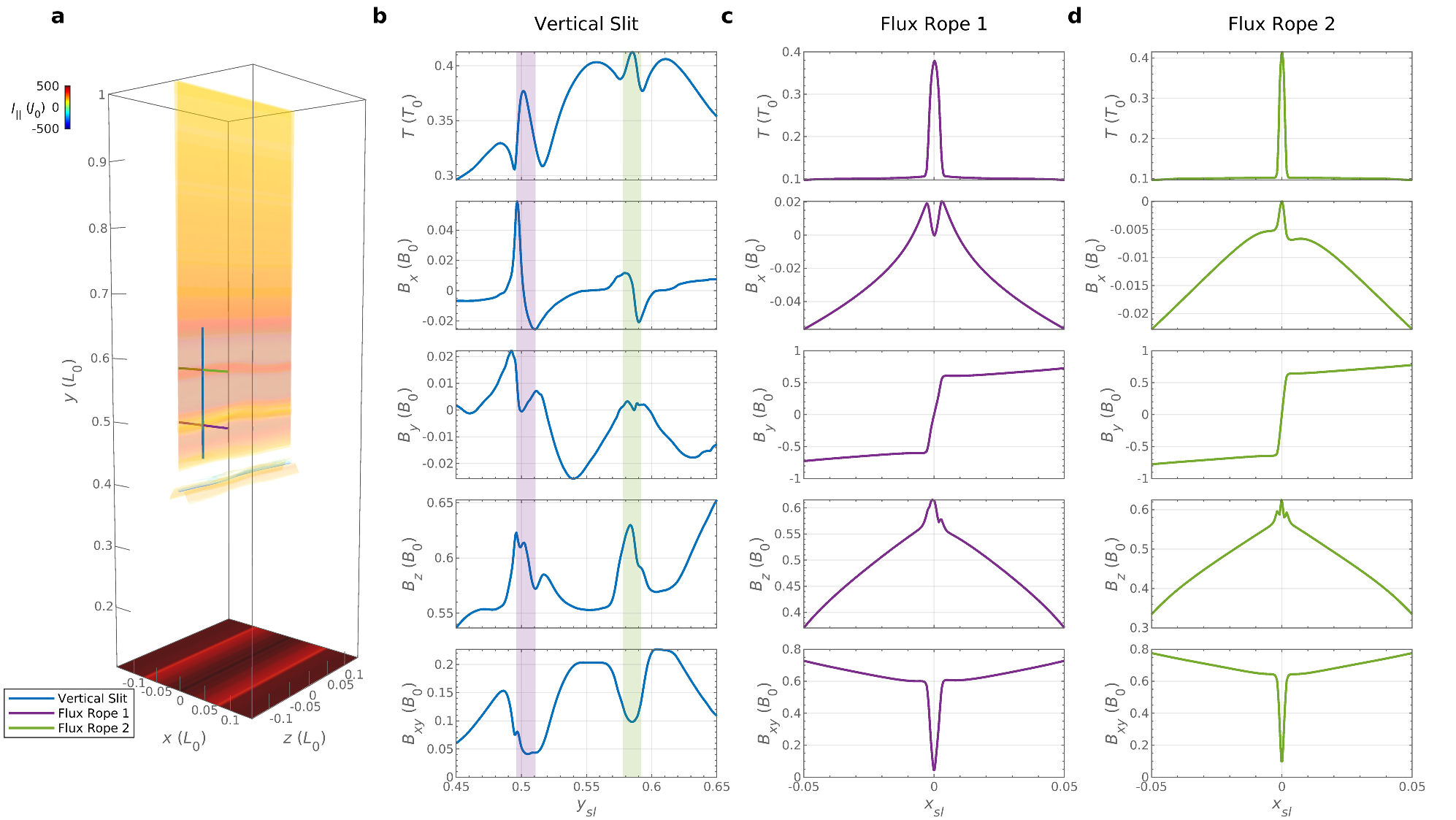}{1\textwidth}{}} 
\caption{Evidence for horizontal flux ropes taking place kink instability at $t=5.7$.
(a) The same content as Fig.\,\ref{fig2}a but at a different moment ($t=5.7$).
The colored lines denote three slits across the axises of two flux ropes.
(b) The profiles of $T$, $B_x$, $B_y$, $B_z$, and $B_{xy}=\sqrt{B_x^2+B_y^2}$ along the vertical slit.
The purple and green shades highlight the regions of flux ropes 1 and 2, respectively.
(c and d) The profiles of $T$, $B_x$, $B_y$, $B_z$, and $B_{xy}$ along the purple and green horizontal slits.
\label{sfig_InitKink}}
\end{figure*}

In the following impulsive stage, $5.7\leq t<7.0$, the reconnection rate increases dramatically and the 3D TMI starts to dominate the evolution of the CS.
At $t=5.7$, the horizontal flux ropes produced by the TMI are characterized by the poloidal radius $r\sim 0.005$, the toroidal length $L_T\sim 0.3$, the toroidal magnetic strength $B_T\approx B_z\sim 0.6$, and the poloidal magnetic strength $B_P\approx B_{xy}\sim 0.1$ (see Fig.\,\ref{sfig_InitKink}).
Their edge safety factors are on the order of $q=2\pi rB_T/L_TB_P\sim 0.6<1$, which satisfies the condition of kink instability \citep{Oz2011,Zhang2021}.
As driven by the kink instability, the previously formed flux ropes are distorted on the third dimension and finally break into pieces, while new flux ropes of various scales also emerge in different orientations.
At the same time, the volume-averaged effective release rate of magnetic energy,
\begin{equation}
    P_M=-\left[\frac{\mathrm{d}}{\mathrm{d}t}\int_{V}\frac{B^{2}}{2}\mathrm{d}V+\oint_{\partial V}\left({\bf E}\times{\bf B}\right)\cdot{\bf n}\mathrm{d}S\right]/\int_V\mathrm{d}V\,,\label{eq:Pm}
\end{equation}
is tripled, reaching a platform of 0.2--0.3 (the blue solid curve in Fig.\,\ref{fig3}a), where, $V$ is the selected region, $\partial V$ denotes its boundary, ${\bf E}=\eta{\bf J}-{\bf u}\times{\bf B}$ is the electric field, ${\bf n}$ is the unit normal vector on the boundary, and $\mathrm{d}S$ denotes the surface element.
The positive value of $P_M$ indicates the release of magnetic energy.
We have performed the normalization by volume for a better comparison between different regions.
As various small-scale structures form, the turbulent magnetic energy with a power spectrum, $E_M^F$, also rises significantly (Fig.\,\ref{fig3}b).
The procedure of evaluating $E_M^F$ is detailed in Appendix \ref{A_TurbMethod}.

After $t\sim 7.0$, the reconnection enters a well-developed turbulent stage with small-scale current structures emerging continuously.
The power spectrum reaches a stable turbulent cascading profile that reflects a global effect including simultaneously the injection, inertial, and dissipation features (Fig.\,\ref{fig3}b).
The power-law index of the inertial range approaches the classical $-5/3$ law as predicted by Kolmogorov and Goldreich-Sridhar (Fig.\,\ref{fig3}b) \citep{Goldreich1995}.
On the other hand, near the end of the simulation, the currents are also concentrated into a few vertically distributed structures forming relatively large-scale quasi-periodic patches with a wavelength of $\sim 0.1$ in the $z$ direction (see Fig.\,\ref{fig2}a), presenting finger-like structures as seen from the face-on view (Fig.\,\ref{fig1}a).
This corresponds to a short concave region near $k\sim 10^{1.5}$ in the spectra, indicating that the turbulent magnetic structures of this scale might be transferred to smaller scales efficiently or experiencing an inversely cascading process via coalescence of smaller structures \citep{Bhat2021}.

Different from the CS region, the development of turbulence at the loop-top region is not fully self-sustained but is driven by the reconnection downflows.
Before $t=7.0$, the average magnetic energy release rate $P_M$ in the loop-top region oscillates near zero (the blue dashed curve in Fig.\,\ref{fig3}a).
Afterward, part of released magnetic energy from the upper CS region starts to accumulate at the loop top, resulting in a negative $P_M$ during $7.0\leq t\leq 7.7$ (Fig.\,\ref{fig3}a).
Since the principal reconnection point in the CS region is about $0.5$ higher than the loop top and the downflow speed is of the order of $1$, the accumulation of magnetic energy is thus delayed by about $0.5$ compared with the fast increase of $P_M$ in the CS region during $t\in\left[6.3,6.8\right]$.
At $t=7$, a strong loop-top termination shock, where electrons are believed to be efficiently accelerated \citep{Chen2015}, appears and gets enlarged in space gradually (see the Movie of Fig.\,\ref{fig2}).
Here, considering that both pressure and magnetic field strength increases across the shock front from upstream to downstream, we approximately locate the positions of the termination shock fronts by the conditions $\nabla B\cdot\nabla p/\left|\nabla p\right|>0$ and $\nabla\cdot{\bf u}\leq D_{tr}$, where the threshold value is set as $D_{tr}=-50$.
After $t=7.7$, the kinetic collision of the reconnection downflows with the loop top starts to explosively develop the turbulence with a spectrum index in the inertial range tending to $-5/3$ (Fig.\,\ref{fig3}c), accompanied by a violent release of the magnetic energy (Fig.\,\ref{fig3}a) and the break of the termination shock (Fig.\,\ref{fig2}).
Compared with that in the CS, the spectrum at the loop top presents a smaller power.
Moreover, the obvious concave region moving towards the larger scale as observed in the CS is not found at the loop top (Fig.\,\ref{fig3}b, c), showing presence of an inversely cascading effect in the CS but absence at the loop top.

\subsection{Development of KHI in the CS}

Our results show that the TMI dominates the initial development of magnetic reconnection similar to \cite{Huang2016}.
However, the KHI is switched on and then causes the formation of relatively large-scale structures in the CS when the shear flows appear.
Here we mainly analyze the KHI that is first triggered by $z$-direction shear flows, which are naturally induced by the evolution of the guide field $B_z$.
Physically, the guide field stems from the horizontal magnetic field over the polarity inverse line (PIL) of solar active regions.
The initial $B_z$ and $B_y$ components form a sheared magnetic field within the CS.
As flux ropes appear, $B_z$ is gradually concentrated towards the center of the CS and then the flux ropes flow upwards and downwards with the outflows.
The movement of flux ropes thus migrates the guide field and causes gradients of $B_z$ on both $x$ and $y$ directions.
At the beginning of the impulsive stage, the system is still approximately symmetric in the $z$ direction.
Thus, we replace $\partial_z$ with 0 and obtain the Lorentz force on the $z$ direction as $\left(\bf{J}\times\bf{B}\right)_z\approx B_y\partial_yB_z+B_x\partial_xB_z$.
With regard to the CS central plane, $B_y$ and $\partial_xB_z$ are asymmetric, while $B_x$ and $\partial_y B_z$ are symmetric, which means that $\left(\bf{J}\times\bf{B}\right)_z$ is asymmetric.
The Lorentz force thus potentially drives the $z$-direction shear flows.

\begin{figure*}[h]
\gridline{\fig{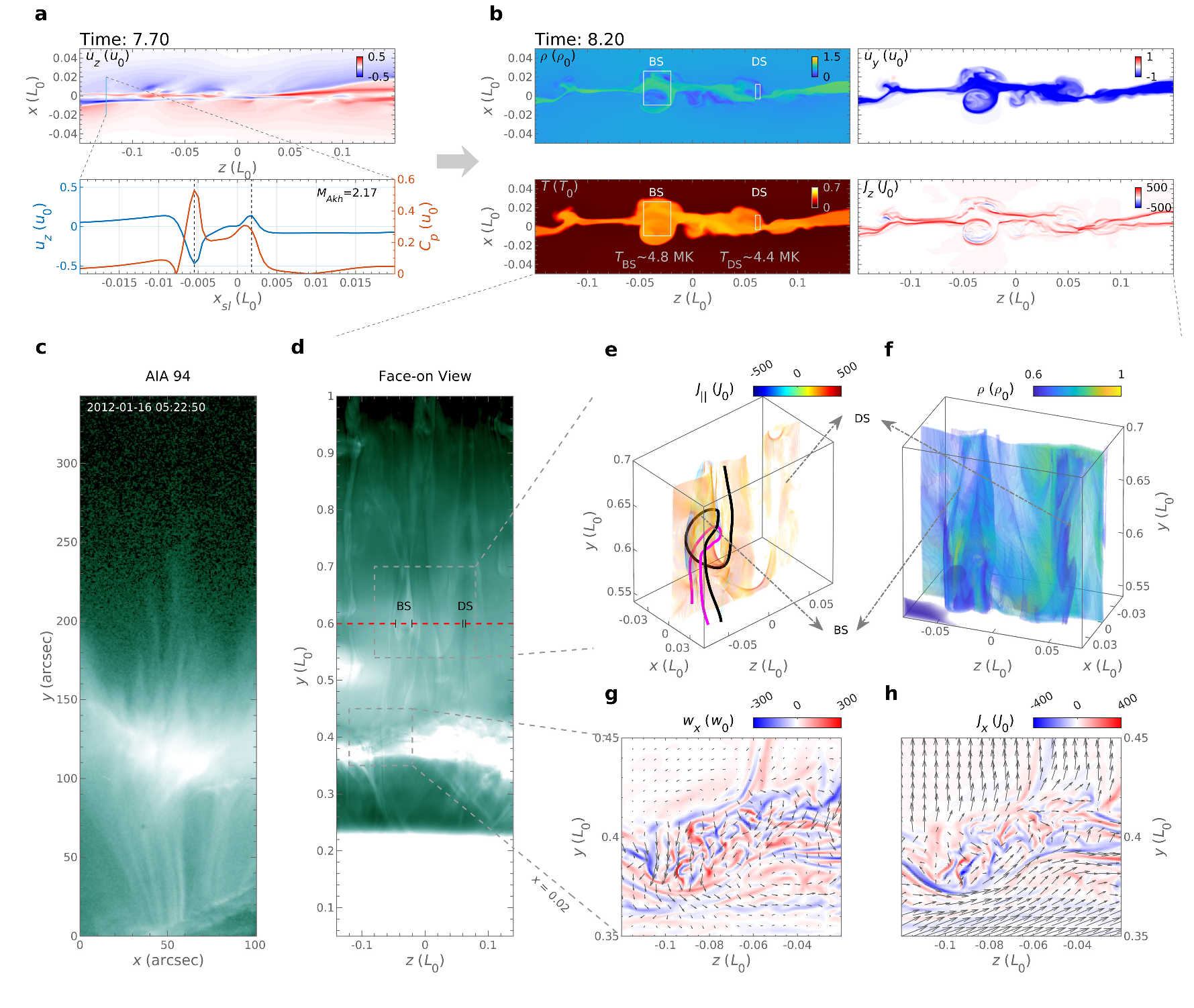}{1\textwidth}{}} 
\caption{Development of KHI and its relation to observations.
(a) The KHI condition of the $z$-direction shear flows in the $x$-$z$ slice on $y=0.6$ at $t=7.7$.
The upper panel shows the profile of $u_z$, in which the cyan line segment is approximately perpendicular to the shear layer.
The lower panel depicts the $u_z$ (the blue curve) and the projected Alfv\'{e}n speed $C_{p}$ (the orange curve) along the segment.
$M_{Akh}$ is the local Alfv\'{e}nic Mach number of the velocity transition across the shear layer enclosed by two dashed lines \citep{Wang2022}.
(b) The distributions of density ($\rho$), vertical velocity ($u_y$), temperature ($T$), and $z$-direction current density ($J_z$) on the slice of (a) but at later moment $t=8.2$.
On the panels of $\rho$ and $T$, we select two regions BS and DS (white boxes) and calculate their average temperature $T_{\mathrm{BS}}$ and $T_{\mathrm{DS}}$, respectively.
(c) The same flare event as Fig.\,\ref{fig1}c observed by the AIA 94\,\AA\ channel.
(d) Synthesized AIA 94\,\AA\ image of the face-on view at $t=8.2$.
The red dashed line denotes the position of the slice in (a) and (b).
Two grey dashed boxes highlight the zoomed-in regions analyzed in panels (e)--(h).
BS and DS mark two typical bright and dark structures, respectively.
(e and f) The 3D rendering of $J_\parallel$ and $\rho$ near BS and DS.
The magenta and black curves in panel (e) depict two magnetic field lines crossing BS.
(g and h) The distributions of out-of-plane vortex $w_x=\left(\nabla\times\bf{u}\right)_x$ and out-of-plane current density $J_x$ on the $y$-$z$ slice at $x=0.02$.
The arrows in (g) and (h) depict the projected velocity field and magnetic field, respectively.
An animation showing the evolution of (a) and (b) from $t=6$ to $8.2$ is available.
\label{fig4}}
\end{figure*}

We use the same method as \cite{Wang2022} to determine the growth condition of the KHI.
The criterion of the KHI in magnetized plasmas is the local Alfv\'{e}nic Mach number of the velocity transition in shear layers \citep{Ryu2000}, which is defined as $M_{Akh}=U_s/C_{p}$, where $U_s$ is the shear velocity, $C_{p}=B_\parallel/\sqrt{\rho}$ is the projected Alfv\'{e}n speed, and $B_\parallel$ is the magnetic strength parallel with the shear flow.
Because the magnetic field perpendicular to the shear flow has little effect on KHI, only the parallel field is considered \citep{Ryu2000}.
To trigger the KHI, $M_{Akh}$ should be larger than $2$ \citep{Jones1997,Ryu2000}.
As shown in Fig.\,\ref{fig4}a, we set a slit approximately perpendicular to the shear flow.
The profiles of the shear speed and $C_p$ are obtained along the slit.
$U_s$ is obtained by $\left|u_{b1}-u_{b2}\right|$, where $u_{b1}$ and $u_{b2}$ are the extreme speeds at the boundaries of the shear layer as marked by the two dashed lines in the lower panel of Fig.\,\ref{fig4}a.
Since $C_{p}$ varies along the slit, we use its averaged value in the shear layer to calculate $M_{Akh}$.

At $t=7.7$, the $z$-direction shear flows initiate the KHI once $M_{Akh}>2$ (Fig.\,\ref{fig4}a) \citep{Wang2022}.
Later on, the KHI gradually grows in the CS region (see the Movie of Fig.\,\ref{fig4}).
At $t=8.2$, the distributions of $\rho$, $u_y$, $T$, and $J_z$ become highly non-uniform and show typical KHI vortex structures corresponding to vertical flux ropes (Fig.\,\ref{fig4}b).
The density and downflow speed are enhanced near the middle plane ($x=0$) and also at the edge of the flux ropes; by contrast, the temperature varies only slightly in the entire CS region (Figs.\,\ref{fig2}b and \ref{fig4}b).
The curled currents, from strong magnetic shear, further promote the reconnection within the flux ropes.

The non-uniform distributions of the temperature and density in the CS are argued to be the essence of alternately bright and dark structures that directly extend toward the higher corona from the flare loop top as often observed, at variance with the interpretation of RTI that locally appears between the TS and flare loop top \citep{Shen2022}.
We select two sub-structures BS and DS, which respectively appear as bright and dark features in the synthetic AIA 94\,\AA\ and XRT images (Figs.\,\ref{fig1}a and \ref{fig4}d), for further inspection.
It is found that the BS encloses a deformed flux rope with a curled current patch wrapped by reconnected magnetic field lines (Fig.\,\ref{fig4}e, f).
The temperature of the BS is mainly distributed near the high-temperature peak of the AIA 94\,\AA\ response function \citep[see][Fig.\,5]{Dere2019}.
The DS is also associated with a flux rope and has a similar temperature with the BS (Fig.\,\ref{fig4}b).
However, it appears dark since its density is lower than that in the BS on average, which causes a lower emission measure integrated along the LOS.

As a consequence of turbulence, the loop-top region is full of small-scale vortexes (Fig.\,\ref{fig4}g, h).
These turbulent structures might result in complex flows \citep{McKenzie2013} and significant non-thermal velocity \citep{Kontar2017a} observed at the flare loop top.
The turbulent structure also causes a number of small-scale CSs, where magnetic energy is further released by the secondary reconnection so as to heat the plasma therein.

\section{Conclusion and Discussions}

In summary, this paper presents a self-consistent 3D simulation of turbulent reconnection within a flare CS.
Starting from a standard CSHKP CS configuration, the reconnection evolves spontaneously and finally reaches a fully turbulent state in both the CS and loop-top regions, exhibiting a number of fine structures, well elucidating the nature of highly dynamic features observed during flares.
Our simulation results achieve a high similarity to observations in both macro and micro scopes.
This is attributed to the fact that our model treats the whole causal chain of the flare reconnection in a more realistic and logical way.
By comparison, in most previous 3D simulations, either ideal configurations or simplified physics were assumed in order to focus on a specific process \citep{Nishida2013,Cecere2015,Edmondson2017,Shibata2023}.
Furthermore, the spatial resolution in our simulation is higher than recent 3D simulations of flare CS reconnection \citep{Shen2022,Ruan2023}.
This is also a vital factor for revealing turbulent structures in the CS region (see Appendix \ref{A_NumConv}).

We set a relatively short $z$ domain compared with the real flare extension along the PIL.
However, the scale of $z$ domain in our simulation is still comparable with the width of the flare loops, which is large enough to reproduce various observational structures that will be investigated in the next paper.
The high spatial resolution ensures the reliability of small-scale reconnection processes caused by small background physical resistivity $\eta_b$ (Appendix \ref{A_NumConv}).
It is worth noting that $\eta_b$ we used is still far larger than that in coronal plasma, where the real magnetic dissipation may extend toward even smaller scales \citep{Dong2022}.
Despite these limitations, our results show a clear physical link from the fundamental 3D turbulent reconnection to the large-scale flare phenomena, largely updating our understanding for flare dynamics at small scales.
The novel phenomena revealed in our model are expected to be able to be compared with high-resolution imaging and spectroscopic observations of flares from next-generation telescopes such as DKIST \citep{Tritschler2015} and MUSE \citep{DePontieu2022}.

\appendix
\restartappendixnumbering

\section{Evaluation of Turbulent Spectra}\label{A_TurbMethod}
We adopt the procedure of calculating turbulence spectra similar to \cite{Huang2016,Beresnyak2017}.
Because the system has approximately translational symmetry along the $z$-direction, the ensemble average of an arbitrary physical quantity $F$ can be estimated by $\bar{F}=\left<F\right>_z$.
The fluctuation part can thus be determined by $\tilde{F}\equiv F-\bar{F}$.
In this work, we analyze the turbulent magnetic energy $E_M=\left|\tilde{\bf B}\right|^2/2$.

To distinguish the physical processes in the CS and Loop-top regions, we use the \textit{Plank-tapper} window function $W^1\left(l,L,\epsilon\right)$ to select the target regions \citep[see][Eq.\,7]{McKechan2010}.
$W^1$ is an even function, which equals $1$ for $\left|l\right|\leq 0.5L\left(1-2\epsilon\right)$ and smoothly reduces to $0$ as $\left|l\right|$ increases to $0.5L$.
The Plank-tapper function can effectively eliminate the unwanted low-frequency components brought by the data truncation at window boundaries.
The 3D window function is constructed from $W^1$ as
\begin{equation}
W^3=W^1\left(x-x_c,L_x,\epsilon\right)W^1\left(y-y_c,L_y,\epsilon\right)W^1\left(z-z_c,L_z,\epsilon\right)\,,
\end{equation}
where, $\epsilon=0.1$, $x_c$ and $L_x$ are respectively the median value and the width of $x$ window, and the definitions are similar for $y_c$, $L_y$ and $z_c$, $L_z$, but for different directions.
The CS window $W^3_{cs}$ is defined by $x\in\left[-0.05,0.05\right]$, $y\in\left[0.45,1\right]$, and $z\in\left[-0.14,0.14\right]$.
Compared with $W^3_{cs}$, the loop-top window $W^3_{lt}$ only varies the $y$ domain as $y\in\left[0.35,0.45\right]$.
The locations of $W^3_{cs}$ and $W^3_{lt}$ have been highlighted by dotted boxes in Fig.\,\ref{fig2}a.

\begin{figure}[h]
\centering
\includegraphics[width=0.8\textwidth]{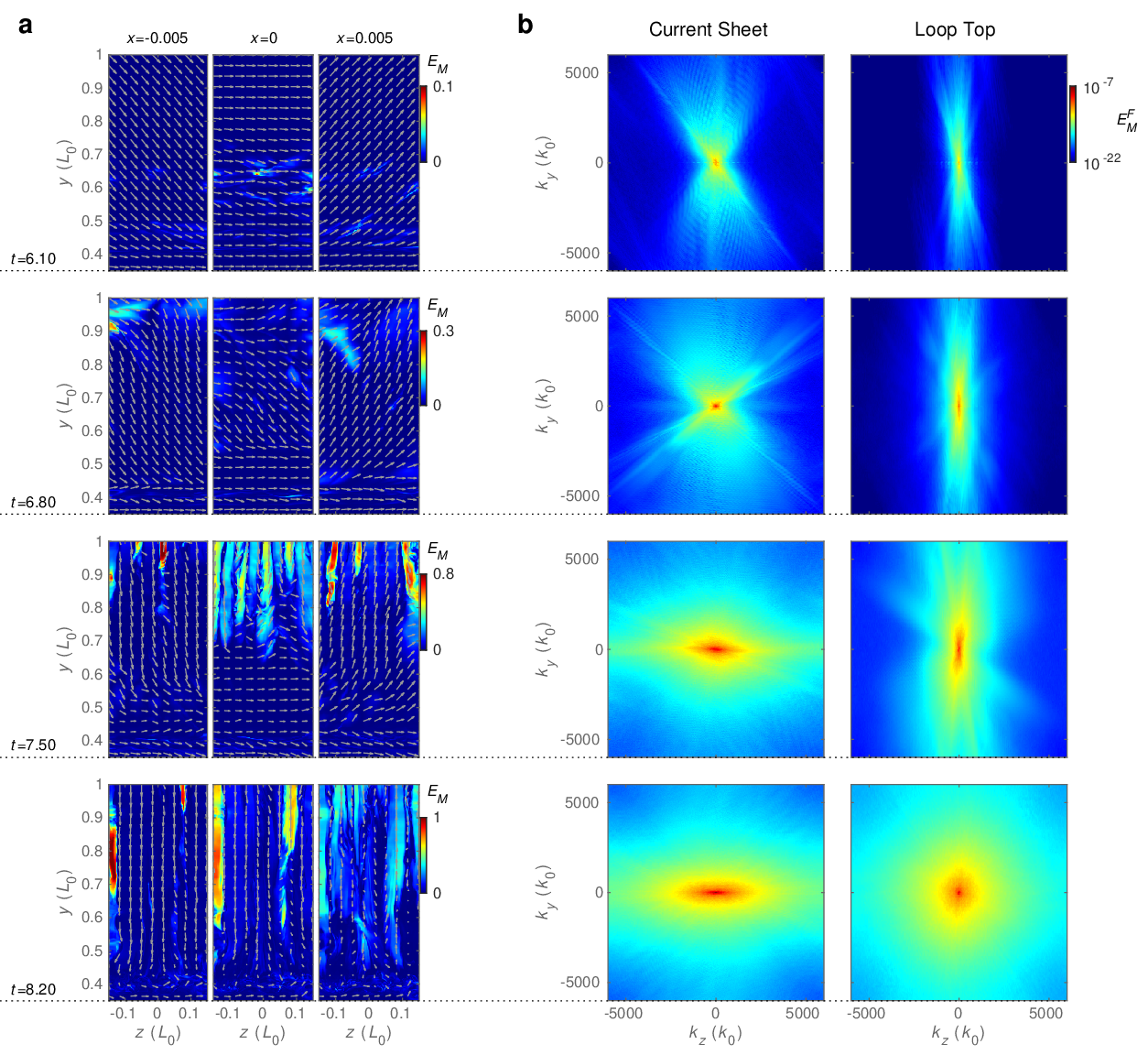}
\caption{Temporal evolutions of turbulent magnetic energy and 2D spectra.
(a) The distributions of turbulent magnetic energy near the CS center.
The three columns show the $E_M$ distribution at $x=-0.005$, $0$, and $0.005$ planes.
The grey arrows depict the in-plane magnetic field.
(b) The distributions of $E^F_M\left(k_z,k_y\right)$ at the CS (left) and the loop-top regions (right).
For both (a) and (b), the four rows correspond to four moments of $t=6.10$, $6.80$, $7.50$, and $8.20$.
\label{sfig_EmTurb2D}}
\end{figure}

There are four steps for evaluating the spectra \citep{Huang2016}.
First, we multiply $E_M$ with $W^3_{cs}$ to select the data in the CS region.
Second, for each $x$ plane, we calculated the 2D power spectra density (PSD), $\hat{E}^F_M\left(x,k_z,k_y\right)$.
Third, because the CS region has significant inhomogeneity in $x$ direction, we average $\hat{E}^F_M$ in $x$ direction and get the mean 2D spectrum $E^F_M\left(k_z,k_y\right)$ (see Fig.\,\ref{sfig_EmTurb2D}b).
Finally, the one-dimensional power spectrum can be obtained by $E^F_M\left(k\right)=\int_0^{2\pi}E'^F_Mk\mathrm{d}\theta$, where $E'^F_M\left(k,\theta\right)$ is the polar coordinate representation of $E^F_M$.

According to Figs.\,\ref{sfig_EmTurb2D}a, the fluctuations of magnetic energy tend to distribute along the local magnetic field.
At the early moment, $t=6.1$, the guide field $B_z$ and background open field $B_y$ form the sheared magnetic field in the CS.
Because $\bf{k}\cdot\bf{B}\simeq0$ is the necessary condition of the TMI \citep{Furth1963,Daughton2011,Huang2016}, the 2D spectrum $E^F_M$ mainly distributes along $y$ direction perpendicular to $B_z$ but also shows a dispersion angle towards $z$ direction (see the 1st. row of Fig.\,\ref{sfig_EmTurb2D}).
As the reconnection develops, the guide field is gradually migrated by outflows and the magnetic field in the CS is dominated by the vertically distributed $B_y$.
As a result, the $E^F_M$ changes toward $z$-direction gradually (see the 2nd. and 3rd. rows of Fig.\,\ref{sfig_EmTurb2D}).
The 2D $E^F_M$ approximately keeps its form during the turbulent stage (see the 4th. row of Fig.\,\ref{sfig_EmTurb2D}).

At the loop-top region, the reconnection downflows inject $B_z$ from the upper CS region.
Before $t=7.7$, the spectra tend to distribute along $y$-direction (see the loop-top spectra at $t=6.1$, $6.8$, and $7.5$ in Fig.\,\ref{sfig_EmTurb2D}b).
Later, the loop-top turbulence grows rapidly and the spectrum forms a relatively isotropic distribution (see the 4th. row of Fig.\,\ref{sfig_EmTurb2D}b).

\section{Numerical Convergence Experiments}\label{A_NumConv}
Besides the high-resolution simulation we analyzed (Case 0), we also perform two additional simulations (Cases 1 and 2) with lower spatial resolutions to test the numerical convergence of our results.
For Cases 1 and 2, all configurations keep unchanged.
The spatial grid lengths in Cases 1 and 2 are respectively two and four times larger than that in Case 0.
In the core regions, the spatial resolutions of Cases 1 and 2 are thus $\Delta L=52\,\mathrm{km}$ and $\Delta L=104\,\mathrm{km}$, respectively.

\begin{figure}[h]
\centering
\includegraphics[width=0.8\textwidth]{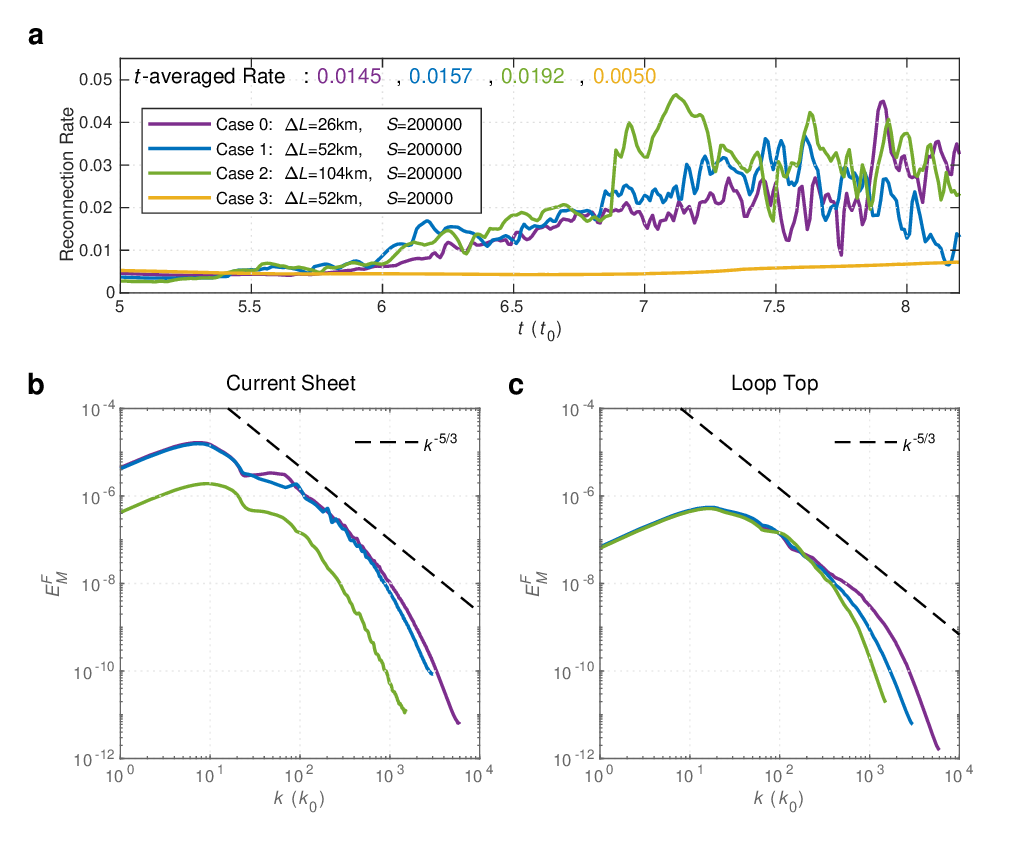}
\caption{Numerical convergences of reconnection rate and turbulent spectra.
(a) The temporal evolution of the reconnection rate for four cases with different resolutions and Lundquist numbers.
The time-averaged reconnection rates are estimated in the time domain $t\in\left[5,8.2\right]$.
(b and c) The spectra of turbulent magnetic energy at $t=8.2$ in the CS (b) and at the loop-top region (c) for four different cases.
\label{sfig_ConvTest_TurbRec}}
\end{figure}

\begin{figure}[h]
\centering
\includegraphics[width=0.8\textwidth]{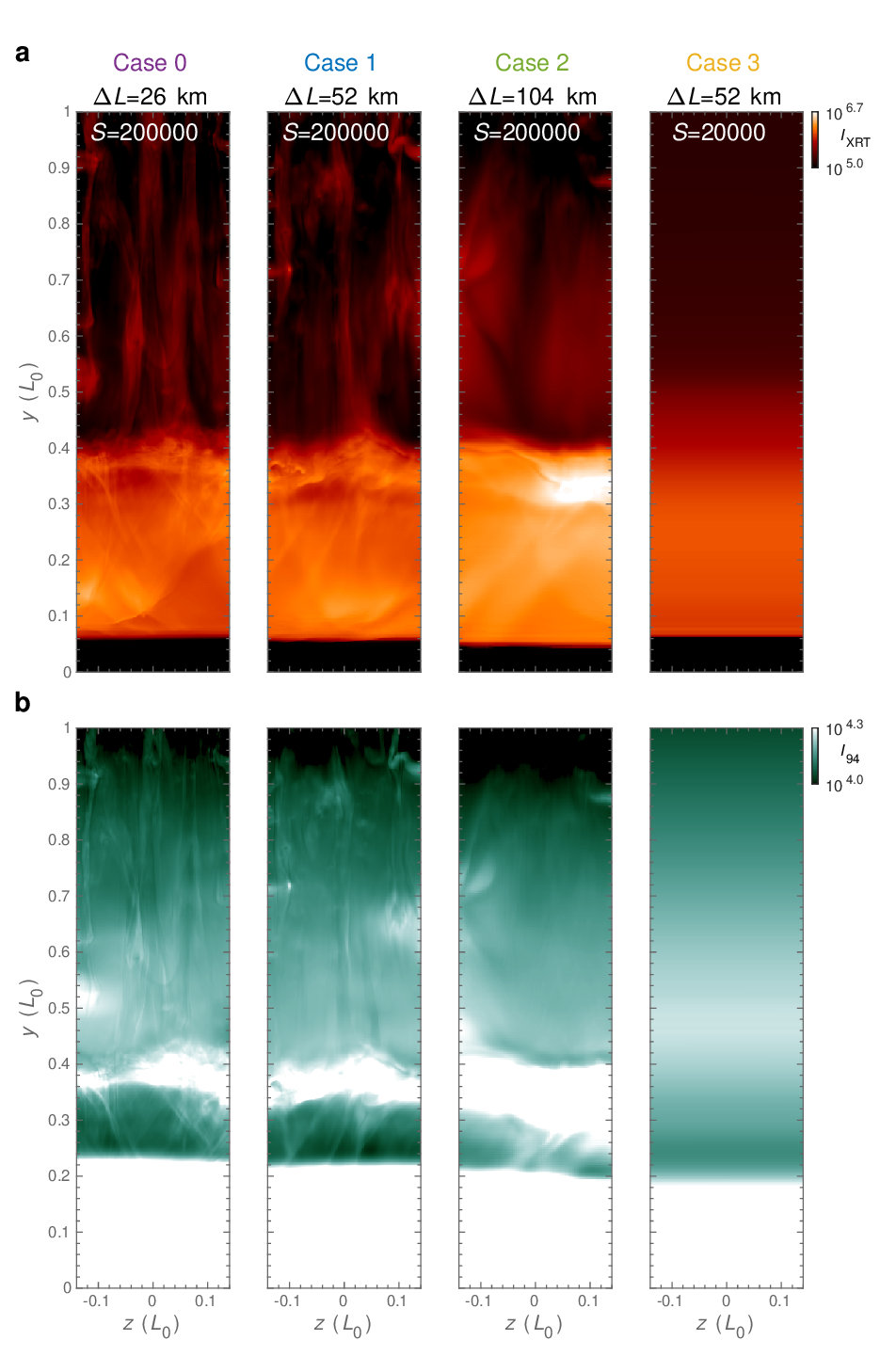}
\caption{Synthesized images of four cases at $\boldsymbol{t}\,\bf{=8.2}$.
(a) The face-on views of the flare CS as observed by the XRT Al-poly/open channel.
(b) The face-on views as observed by the AIA 94\,\AA\ channel.
\label{sfig_ConvTest_Obs}}
\end{figure}

Compared with Case 0, the reconnection rate of Case 1 shows a similar evolution trend and time-averaged value as Case 0 (Fig.\,\ref{sfig_ConvTest_TurbRec}a).
Meanwhile, the turbulent energy spectra of Case 1 exhibit injection and inertial ranges similar to Case 0, even though the small-scale dissipation range cannot be well resolved (Fig.\,\ref{sfig_ConvTest_TurbRec}b and c).
Furthermore, the synthesized images from the Case 1 data also capture the finger-like structures (Fig.\,\ref{sfig_ConvTest_Obs}).
Therefore, we conclude that the differences in the spatial resolutions and thus the numerical resistivity between Case 0 and 1 have a limited influence on our main results.

However, Case 2 outputs significantly different results.
Because the increase of $\Delta L$ results in a larger numerical resistivity, the reconnection rate of Case 2 is greater than that of Case 0 (Fig.\,\ref{sfig_ConvTest_TurbRec}a).
More importantly, in the CS region, the turbulent energy spectra cannot form inertial ranges, which implies that the turbulence is not sufficiently developed (Fig.\,\ref{sfig_ConvTest_TurbRec}b).
Correspondingly, in synthesized images of Case 2, the finger-like structures are invisible above the loop top (Fig.\,\ref{sfig_ConvTest_Obs}).
However, at the loop-top region, we still obtain the inertial region of the turbulence spectra with similar spectrum indexes (Fig.\,\ref{sfig_ConvTest_TurbRec}c).
The reason could be that the loop-top turbulence is mainly dependent on the reconnection downflows, which can induce various instabilities driving the turbulence.
Comparatively speaking, the difference in numerical resistivity might have insignificant influences on the development of the loop-top turbulence.

In addition, we test the influences of physical resistivity by Case 3 which sets the same spatial resolution with Case 1 but a much larger background resistivity $\eta_b=5\times 10^{-5}$ corresponding to a Lundquist number $S=2\times 10^4$ (see Figs.\,\ref{sfig_ConvTest_TurbRec}a and \ref{sfig_ConvTest_Obs}).
Because $S$ is too small to initiate the TMI, the reconnection in Case 3 presents a typical Sweet-Parker pattern without the break of CS or the growth of 3D turbulence.
Its reconnection rate, as shown by the yellow curve in Fig.\,\ref{sfig_ConvTest_TurbRec}a, is on the order of $S^{-1/2}$, consistent with the Sweet-Parker rate \citep[see][]{Huang2010}.
Consequently, the synthesized face-on images are uniform on the $z$-direction (see Fig.\,\ref{sfig_ConvTest_Obs}). 
This test implies that a large Lundquist-number environment is vital to the development of 3D turbulent reconnection.

\section*{Acknowledgments}
We would like to acknowledge fruitful discussions with X. Bai and J. Chen.
High-performance computing resources supporting this work were provided by National Supercomputing Center in Jinan.
This work was supported by the National Key R\&D Program of China under grant 2021YFA1600504, by the National Natural Science Foundation of China under grants 12127901, and by the Shandong Provincial Natural Science Foundation under grant ZR2021QA083.


\end{CJK}
\end{document}